\documentclass[letterpaper,times]{sae}

\usepackage[svgnames]{xcolor}
\definecolor{SAEblue}{RGB}{1,160,233}
\raggedright

\usepackage[justification=justified, singlelinecheck=false]{caption}  
\DeclareCaptionFont{eightpt}{\fontsize{8pt}{11pt}\selectfont #1}
\usepackage{lineno}
\usepackage[utf8]{inputenc}
\usepackage{cite}
\usepackage{array}
\newcolumntype{L}[1]{>{\raggedright\let\newline\\\arraybackslash\hspace{0pt}}p{#1}}
\newcolumntype{C}[1]{>{\centering\let\newline\\\arraybackslash\hspace{0pt}}p{#1}}
\newcolumntype{R}[1]{>{\raggedleft\let\newline\\\arraybackslash\hspace{0pt}}p{#1}}

\usepackage{graphicx}
\usepackage{multirow}
\usepackage{amsmath}
\usepackage{url} 
\newcommand{\ignore}[1]{}

\usepackage{nameref}
\usepackage{booktabs}

\makeatletter
\def\@seccntformat#1{%
  \expandafter\csname c@#1\endcsname\c@section
  }
\makeatother

\usepackage{subfigure}
\usepackage{multimedia}
\usepackage{siunitx}
\usepackage{float}

\PaperTitle{Eco-Driving Control for Electric Vehicles with Multi-Speed Transmission: Optimizing Vehicle Speed and Powertrain Operation in Dynamic Environments}
\usepackage{calc} 
\usepackage{ragged2e}
\makeatletter 
\renewcommand\@biblabel[1]{#1. } 
\makeatother

\newsavebox{\authorbox} 
\savebox{\authorbox}{}

\makeatletter
\newcommand{\AddAuthor}[2]{%
  \savebox{\authorbox}{%
    \parbox[t]{\textwidth}{%
      \flushright%
      \usebox{\authorbox}\\%
      \fontsize{12}{14}\bfseries #1\\%
      \fontsize{10}{11}\mdseries\itshape #2\\[2ex]%
    }%
  }%
}

\def\@maketitle{%
  \null
  \parbox[t][45mm][t]{\textwidth}{
    \begin{flushright}%
      \fontsize{14}{18}\selectfont%
      \usebox{\numberbox}\\
      \usebox{\titlebox}\\[2ex]
      \usebox{\authorbox} 
    \end{flushright}%
    \vfill
    \begin{flushleft}%
        \raisebox{2cm}{\usebox{\saecopyrightbox}}\\%
    \end{flushleft}}%
}
\makeatother

\AddAuthor{Suiyi He}{Department of Mechanical Engineering, University of Minnesota, Minneapolis, MN, USA}
\AddAuthor{Zongxuan Sun}{Department of Mechanical Engineering, University of Minnesota, Minneapolis, MN, USA}

\begin{document}
\maketitle


Connected and Autonomous Electric Vehicles\\
Multi-Speed Transmission\\
Traffic Prediction

\justifying

\section{Abstract}
This article presents an eco-driving algorithm for electric vehicles featuring multi-speed transmissions.
The proposed controller is formulated as a co-optimization problem, simultaneously optimizing both vehicle longitudinal speed and powertrain operation to maximize energy efficiency.
Constraints derived from a connected vehicle based traffic prediction algorithm are used to ensure traffic safety and smooth traffic flow in dynamic environments with multiple signalized intersections and mixed traffic.
By simplifying the complex, nonlinear mixed integer problem, the proposed controller achieves computational efficiency, enabling real-time implementation.
To evaluate its performance, traffic scenarios from both Simulation of Urban MObility (SUMO) and real-world road tests are employed.
The results demonstrate a notable reduction in energy consumption by up to 11.36\% over an \SI{18}{\km} drive.

\section{Introduction} \label{sec:ev-eco-introduction}

Global efforts to cut greenhouse gas emissions have made vehicle energy efficiency a top priority. In 2023, the U.S. transportation sector consumed nearly 28 quadrillion British thermal units of energy, with cars and trucks accounting for 80\% of that total~\cite{energy2024}. As demand is projected to rise, improving vehicle efficiency remains a critical path toward global sustainability.

Electric vehicles (EVs) are key to sustainable transport, but maximizing their efficiency is essential for range and cost-effectiveness \cite{muratori2021rise}. 
{To design an energy-efficient EV powertrain, the authors in \cite{gao2025multi} propose a method that optimizes the powertrain sizing, energy management and eco-driving simultaneously.
Alternatively, when the powertrain architecture is fixed, one promising solution lies in developing various types of eco-driving controllers, where both model-based and learning-based methods have shown the capability to reduce overall energy consumption \cite{shao2019optimal, naeem2022eco,hamednia2023optimal,rabinowitz2023real, fan2024deep, lee2020model}.}
{In particular, with the emergence of {connected and autonomous vehicles (CAVs)}, real-time traffic information can be used to design these controllers \cite{coppola2022eco, li2024review, zhao2025design}.}
With cellular vehicle-to-everything (C-V2X) or dedicated short range communication (DSRC) technology, CAVs are capable of accessing real-time data.
This includes receiving Signal Phase and Timing (SPaT) from traffic signals through Vehicle-to-Infrastructure (V2I) communication and obtaining speed and position data from nearby CAVs via Vehicle-to-Vehicle (V2V) communication.
Additionally, onboard sensors of the ego vehicle can also detect and interpret traffic information related to nearby human-driven vehicles (HVs).
This allows the eco-driving system to optimize energy consumption while maintaining safety and smooth traffic flow~\cite{shao2020vehicle, shao2021energy, sun2022energy}.

While EVs traditionally use single-speed transmissions for simplicity, there is a growing shift toward multi-speed transmissions.
However, recent advancements in EV technology have driven renewed interest in introducing multi-speed transmissions to electric powertrains \cite{ruan2016comparative,ahssan2018electric, gao2022topology, lacock2023electric}.
This transition can significantly improve energy efficiency, extend driving range, and enhance acceleration compared to single-speed systems.

{Improving the multi-speed transmission's design for EVs has been the focus of multiple research efforts \cite{mazali2022review}.}
In \cite{ruan2017investigation}, a novel energy storage system combining a multi-speed transmission and supercapacitor is proposed to improve EVs' driving range and battery life.
The mechanical design of a two-speed transmission and its corresponding control algorithm are presented in~\cite{sorniotti2012analysis, saini2024innovating}.
Authors in \cite{liu2023study} present a novel continuously variable transmission (CVT) design specific for EVs.
Other works, such as~\cite{gao2015gear, zhu2015gear, fang2016design,huang2020optimal, nguyen2021optimization, zhou2025design}, aim to optimize shift schedules to improve both dynamic performance and energy efficiency.
Dynamic programming (DP) is used in~\cite{han2019optimized} to achieve optimal designs, including the number of gears, gear ratios, and shift schedules for multi-speed transmissions. Meanwhile, learning-based approaches in~\cite{kwon2023optimization} refine the design of two-speed transmissions, accounting for the efficiency of different powertrain components.
Similarly, such methods are also used in \cite{kim2024driving} to optimize the gear-ratio design of a two-speed transmission for electric vehicles considering battery state of charge consumption and vehicle's acceleration performance.
Researchers have also dedicated efforts to improving gear shift quality of EVs' multi-speed transmissions by considering various factors.
Control algorithms to enhance shift smoothness are explored in~\cite{qi2017analysis, hong2016shift, chai2019compound} while work in \cite{yang2023optimal} aims to reduce energy loss during the gear shift process.
To consider these factors in a unique optimization framework,~\cite{tian2020optimal} introduces a controller designed to minimize vehicle jerk and friction work of a two-speed transmission for EVs.

{However, because traffic conditions are highly dynamic, these pre-calculated shift schedules might not provide optimal energy efficiency in real-time eco-driving applications.
To improve energy efficiency in eco-driving scenarios, in addition to above architecture optimization, several studies have focused on optimizing gear selection in real time.}
For instance, by predicting the torque demand of human drivers, authors in \cite{guo2016online, guo2017line} optimize the shift schedule of EV multi-speed transmissions online to reduce energy consumption.
Similarly, \cite{liao2021eco} presents a DP-based algorithm to improve EV energy efficiency in the context of a continuously variable transmission (CVT).
In \cite{sun2023coordinated}, a model predictive control (MPC) based algorithm optimizes vehicle speed and powertrain operation for EVs equipped with a two-speed transmission by approximating the discrete gear positions using a continuous function.
{Similarly, the authors in \cite{koch2021eco} use soft constraints in the cost function to enable gear shifts in a two-speed transmission while assuming exact knowledge of the preceding vehicle's future movement.}
To address the computational challenges of mixed-integer programming for discrete gear positions, researchers use the hierarchical control structure~\cite{han2020hierarchical} and continuous relaxation technique \cite{li2021coordinated} to simplify the optimization problem.
Despite their potential, these approaches exhibit notable limitations. 
Both these approaches assume that the mechanical braking forces are always zero, considering only forces from motor regeneration, which fails to accurately represent the real powertrain dynamics of an EV.
Furthermore, the proposed hierarchical controller in \cite{han2020hierarchical} divides the problem into several sub-tasks and cannot guarantee the optimality of the entire powertrain system.
Similarly, while the optimization problem in~\cite{li2021coordinated} is simplified, it still lacks guarantees for real-time solvability.
These drawbacks limit the effectiveness of these algorithms in improving energy efficiency of eco-driving controller.

{While real-time gear optimization is well-established for internal combustion engine (ICE) vehicles \cite{li2021online, shao2020vehicle, bentaleb2024gear}, a significant gap remains for multi-speed Electric Vehicles (EVs). Current EV methods often struggle with computational complexity or sacrifice optimality.
To address this, this work develops a real-time co-optimization eco-driving controller for Connected and Autonomous EVs (CAEVs). The key advantages of this algorithm include:
\begin{itemize}
    \item \textbf{Simultaneous Co-Optimization:} Unlike hierarchical methods \cite{han2020hierarchical}, this controller concurrently optimizes vehicle speed, motor operation, and discrete gear shift maneuvers for a multi-speed transmission.
    \item \textbf{High-Fidelity Modeling:} The formulation incorporates a comprehensive powertrain model, including battery dynamics, motor constraints, and—unlike previous studies—realistic regenerative and mechanical braking interactions.
    \item \textbf{Superior Stability and Efficiency:} Compared to data-heavy reinforcement learning (RL) \cite{lee2020model} or computationally intensive Nonlinear MPC (NMPC) \cite{shao2019ASME}, this model-based approach is more data-efficient and ensures real-time solvability.
    \item \textbf{Safety-Critical Constraints:} By leveraging V2X traffic predictions, the controller ensures car-following safety and smooth traffic flow.
\end{itemize}}
{To validate the effectiveness of the proposed eco-driving control design, 
traffic scenarios from both numerical simulations and real-world road tests are used.
The results show that the proposed algorithm achieves up to an 11.36\% energy benefit over an \SI{18}{\km} drive compared to the controller for the EV with single-speed transmissions.}

Following subsections are included in this paper: 
First, we briefly introduce the traffic prediction algorithm used in this work.
Next, we show the proposed eco-driving co-optimization controller for EVs with multi-speed transmissions.
To validate the performance of our proposed controller, traffic scenarios from both numerical simulations and road tests are used to test the algorithm. 
The final section shows the key conclusions from this work.

\section{Traffic Prediction} \label{sec:ev-eco-background}
\begin{figure}
    \centering   
    \includegraphics[width=0.95\linewidth]{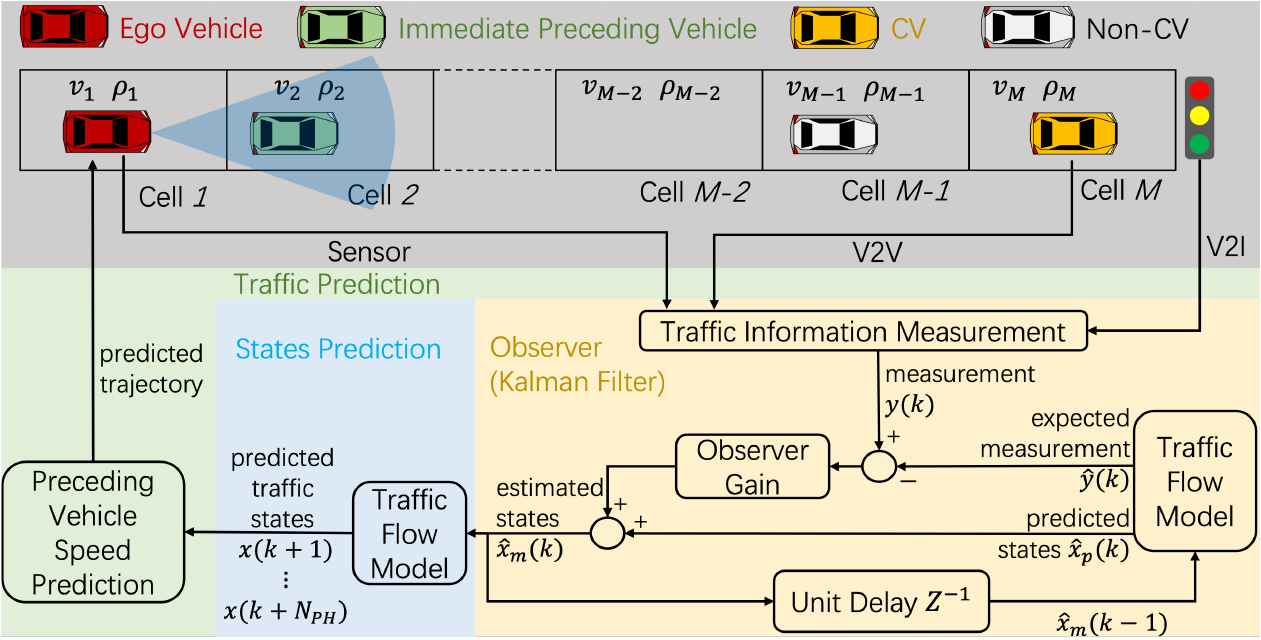}
    \caption{Traffic prediction framework used in this work.
    The red car represents the ego vehicle that operates using the proposed algorithm.
    Yellow vehicles indicate CVs, while gray vehicles represent non-CVs.
    Using the real-time traffic information obtained through communications and sensors, the algorithm predicts the longitudinal movement of the vehicle directly ahead.}
    \label{fig:ev-eco-prediction-structure}
\end{figure}

To improve the ego vehicle's energy efficiency, the eco-driving controller must optimize the ego vehicle's longitudinal operation by considering its immediate preceding vehicle's predicted trajectory.
The ego vehicle must react to the actions of the vehicle ahead to {maintain a safe following distance} and uphold overall traffic efficiency.
This section outlines the traffic prediction approach integrated into the eco-driving controller. Specifically, the method proposed by \cite{shao2020eco} is employed to forecast traffic conditions along the ego vehicle’s lane over the prediction horizon. To account for multi-lane traffic dynamics, the framework can be extended using the approach from \cite{he2023real,he2025connectivity}, which incorporates lane-change behavior to enhance prediction accuracy.
{As shown in \cite{he2025connectivity, sun2021traffic}, this method provides reliable traffic prediction output even when only the immediate preceding vehicle's information is available, and it remains effective across various connected vehicle penetration rates. Detailed analyses of the algorithm's uncertainty and root mean square errors (RMSE) are provided in these existing studies.}

The prediction algorithm used for this work is presented in Figure~\ref{fig:ev-eco-prediction-structure}.
First, by using an unscented Kalman filter (UKF)~\cite{wan2000unscented}, it estimates the driving conditions in terms of traffic speed and traffic density of the road section ahead of the immediate preceding vehicle.
Real-time SPaT data, along with the speed and position data from the ego vehicle (via GPS), the immediate preceding vehicle (via onboard sensors), and connected vehicles (via V2V communication when within the communication range), serves as partial measurements.
Then, the algorithm propagates the traffic flow model to obtain future traffic conditions along the prediction horizon.
Because a vehicle's speed is strongly influenced by the surrounding traffic flow speed, the algorithm is able to estimate the preceding vehicle's longitudinal velocity and trajectory with reasonable accuracy.

The following discretized second-order Payne Whitham (PW) model~\cite{wang2005real} is used in the algorithm:
\begin{subequations}
\setlength{\abovedisplayskip}{3pt}
    \setlength{\belowdisplayskip}{3pt}
\begin{align}
    \rho_{j}(k+1) & = \rho_{j}(k) - \frac{dt}{d x} \left[ \rho_{j}(k)v_{j}(k) - \rho_{j-1}(k)v_{j-1}(k) \right] \nonumber\\
     &\hskip80pt + \omega_{j}(k), \label{eq:ev-eco-rho}\\
     v_{j}(k+1) &= v_{j}(k) - \frac{dt}{d{x}}v_{j}(k)[v_{j}(k)-v_{j-1}(k)] \nonumber\\
     &\hskip30pt + \underbrace{dt \cdot \frac{[{V}_{e}(\rho_{j}(k))-v_{j}(k)]}{\tau}}_{\text{Speed adaptation term}} \nonumber \\
     &\hskip10pt - \underbrace{\frac{dt}{d{x}} \cdot \frac{c_{0}^{2} \cdot [\rho_{j+1}(k)-\rho_{j}(k)]}{\rho_{j}(k)+\epsilon} }_{\text{Traffic pressure term}}+ \xi_{j}(k),\label{eq:ev-eco-v}
\end{align}\label{eq:ev-eco-pw}
\end{subequations} \hfill \break
where \eqref{eq:ev-eco-rho} and~\eqref{eq:ev-eco-v} show density and speed evolutions of the traffic flow, respectively.
In these equations, $k$ denotes the discrete time index, $dt$ is the time step length, and $dx$ shows the length of each cell.
The variable $j$ refers to the index of the cell.
A small constant $\epsilon$ is included to avoid division by zero.
To account for model uncertainties, the terms $\omega_{j}(k)$ and $\xi_{j}(k)$ are introduced and are assumed to follow Gaussian distributions.
The parameter $c_{0}$ quantifies traffic pressure, while $\tau$ defines the rate at which vehicles adjust their speed toward equilibrium.
The equilibrium speed in cell $j$, denoted as ${V}_e(\rho_{j}(k))$, is determined using a triangular fundamental diagram that relates speed and density:
\begin{equation}
	V_{e}(\rho_{j}(k)) =
	\begin{cases}
		v_{0}, & 0 \leq \rho_{j}(k) < \rho_{c}    \\
		c \left(\dfrac{\rho_{\text{jam}}}{\rho_{j}(k)} - 1\right), & \rho_{c} \leq \rho_{j}(k) \leq \rho_{\text{jam}}
	\end{cases}   \label{eq:orignial-equilibrium}
\end{equation}
with
\begin{equation}
    \rho_{c} = \frac{\rho_{\text{jam}}}{v_{0}/c + 1}    \label{eq:ev-eco-rho-c}
\end{equation}
where $v_{0}$ represents the free-flow speed, while $c$ is the gradient of the traffic density drop under congested traffic conditions.
$\rho_{\text{jam}}$ corresponds to the traffic jam density, and the critical density $\rho_{c}$ is defined by \eqref{eq:ev-eco-rho-c}.

{To capture the influence of signalized intersections on traffic flow dynamics, the algorithm sets the traffic speed $v_{s}(k)$ to zero in~\eqref{eq:ev-eco-v} during red lights at the cell $s$, which represents the location of the traffic signal.}


Then, the following linear interpolation is applied to compute the driving speed of the $i$-th vehicle by blending the traffic speeds of the two cells that are adjacent to its location on the driving lane:
\begin{equation}
    y_i(k) = (1-\alpha_i(k))v_{j_\text{adj}}(k)+\alpha_i (k)v_{j_\text{adj}+1}(k) + \phi_i(k) \label{eq:spd-measure}
\end{equation}
where $j_\text{adj}$ denotes the index of the most recently passed cell by the vehicle.
The interpolation weight, $\alpha_i(k)$, is calculated as $ \alpha_i(k)=d_i/dx - j_\text{adj}$.
Here, $d_i$ is the longitudinal location of the $i$-th CV.
The term $\phi_{i}(k)$ accounts for measurement noise, which is assumed to follow a Gaussian distribution.
This enables the longitudinal speed and location of CVs to serve as partial measurements for the UKF.
Once the estimated traffic states are obtained, the algorithm calculates the future traffic speed and the predicted longitudinal movement of the vehicle directly along the prediction horizon.

\section{Optimizing Vehicle Speed and Powertrain Operation Simultaneously}\label{sec:ev-eco-spd-optimization}

Here, we present the design of a co-optimization problem for vehicle longitudinal speed and powertrain operation in EVs equipped with multi-speed transmissions.
This study considers an EV equipped with a single motor located on the rear axle, which delivers power to the rear wheels through a multi-speed transmission.
In the following results demonstration, all parameter values are sourced from Autonomie~\cite{moswd2020autonomie}.

\subsection{System Dynamics} \label{sec:ev-eco-system-dynamics}
This subsection introduces the longitudinal dynamics of the vehicle and the battery dynamics model used in the co-optimization problem.
The corresponding equations are shown below:

\begin{equation}
    \begin{bmatrix} \dot{{d}}(t) \\ \dot{{v}}(t) \\ \dot{SOC}(t) \end{bmatrix} = \begin{bmatrix} {v}(t) \\ {a}(t) \\ -\dfrac{{I}_b(t)}{Q_b}  \end{bmatrix}
\end{equation}
where ${d(t)}$ is the vehicle's longitudinal location at time $t$.
$v(t)$ and $a(t)$ describe its longitudinal speed and acceleration, respectively.
The term $SOC(t)$ refers to the battery's state of charge, $I_b(t)$ indicates its output current, and $Q_b$ presents the total battery capacity.

\subsection{Powertrain Model} \label{sec:ev-eco-powertrain-model}
To illustrate the ego EV's powertrain operation with the inclusion of a multi-speed transmission, the following powertrain model is used~\cite{sun2014design}:
\begin{subequations}\label{eq:ev-eco-prediction-powertrain}
\begin{align}
    w_m(t) &= r_g(t)   \cdot {v}(t) \label{eq:ev-eco-powertrain-model} \\
    {T}_m(t) r_g(t) &= m\cdot {a}(t) + {F}_b(t)+{f}_{\varphi}(t) + C_{wind} {v}^2(t)  \label{eq:ev-eco-vehicle-model}
\end{align}
\end{subequations}
where $w_m(t)$ represents rotational speed of the motor, while $r_g(t)$ captures the lumped ratio that considers the final drive wheel radius, and transmission gear ratios.
The term $T_m(t)$ represents the motor torque.
The total resistance due to rolling and road grade is given by ${f}_{\varphi}(t) = mg \cdot \sin\left[\varphi({d}(t))\right] + \mu mg \cdot \cos\left[\varphi({d}(t))\right] $, where $\varphi(d(t))$ denotes the slope angle at the vehicle's position-typically available from navigation data.
The parameters $\mu$, $m$, and $g$ correspond to the rolling resistance factor, vehicle mass, and gravitational acceleration, respectively.
Air drag coefficient is captured by the constant $C_{wind}$, and $F_b(t)$ represents the (positive) hydraulic braking force.

\subsection{Battery Model}\label{sec:ev-eco-battery-model}

In this work, the battery is modeled using an internal resistance model, whose output power is defined by the following equation:
\begin{equation}
    P_b(t) = n_s {V}_{oc}(SOC) \cdot I_b(t) - I^2_b(t)\cdot \dfrac{n_s}{n_p} R_b(SOC)\label{eq:ev-eco-battery-model}
\end{equation}
where $P_b(t)$ denotes the battery output power;
he terms $V_{oc}(SOC)$ and $R_b(SOC)$ correspond to the open-circuit voltage and internal resistance of a single cell, both of which depend on the current state of charge ($SOC$).
The battery pack is configured with $n_p$ cells connected in parallel per group and $n_s$ such groups connected in series.
{Although $V_{oc}(SOC)$ and $R_b(SOC)$ vary with $SOC$, as observed in the result section, their changes are relatively minor over the prediction horizon. Similar to previous work \cite{shao2019ASME}, to reduce computational complexity, the algorithm updates their values only once at the start of each optimization cycle.}


\subsection{Cost Function Design}
\noindent
The goal of the optimization problem is defined by the following objective function:
\begin{eqnarray} \label{eq:ev-eco-objective-function}
    J = w_0\left({d}(t_f)-{d}_z\right)^2 + \int^{t_f}_{t_0}\left[w_1 {P}_b(t) + w_2\cdot {a}^2(t)\right] \mathrm{d}t
\end{eqnarray}
where $\left[t_0, t_f\right]$ defines the time interval over which the optimization is performed.
The term $w_0$ is a positive weighing factor used to penalize the deviations between the ego vehicle's predicted terminal location and the desired terminal location $d_z$;
$w_1$ and $w_2$ are positive weighting factors describing energy consumption and comfort, respectively.

In this work, the desired terminal location $d_z$ is defined as a position located behind the predicted position of the vehicle directly ahead, and is calculated using the following equation:

\begin{equation}
    d_z = d_\text{lead}(t_f) - 2\cdot h_{min}v_\text{lead}(t_f),
\end{equation}
where $d_\text{lead}(t_f)$ and $v_\text{lead}(t_f)$ are the immediate preceding vehicle's predicted location and speed at terminal horizon $t_f$, respectively.
$h_{min}$ denotes the time headway is set to \SI{1.5}{\second}.
This formulation provides a reference terminal location for the ego vehicle, which serves as a guide for the eco-driving controller.


Two elements make up the battery output power.
\begin{equation}
    P_b(t) = P_{aux}(t) + P_{drv}(t)/\eta_\text{gear}
\end{equation}
where $P_{drv}(t)$ is the drive power, as defined by~\eqref{eq:ev-eco-drive-power-model}.
$\eta_\text{gear}$ indicates the powertrain efficiency.
{The term $P_{aux}(t)$ denotes the auxiliary power, such as the load consumed by the air conditioning system. Since this power demand is not optimized by the algorithm and is controlled by the vehicle's occupants, it is assumed to remain constant across the entire optimization horizon to simplify computation.}

In this context, $P_{drv}(t)$ denotes the drive power, as defined in Equation~\eqref{eq:ev-eco-drive-power-model}; $\eta_\text{gear}$ represents the efficiency of the powertrain. The term $P_{aux}(t)$ accounts for auxiliary loads, including components like the air conditioning system and the 12V low-voltage battery. To simplify the problem, this part is assumed to be unchanged throughout the optimization horizon.

{The drive power $P_\text{drv}(t)$ propels the vehicle.}
Since the motor efficiency generally depends on its operating conditions-specifically the motor speed $w_m(t)$ and torque $T_m(t)$-different equations are applied to capture the relationship between $P_\text{drv}(t)$ and the motor's behavior, {depending on whether it is functioning in propulsion or regeneration mode}:
\begin{eqnarray}
&~&\hskip-20pt P_\text{drv}(t) = \nonumber\\
    &~&\hskip-20pt\begin{cases}
	    \dfrac{w_m\left(t \right)T_m\left(t\right)}{\eta_m\left(T_m\left(t\right),w_m\left(t\right)\right) },~ T_m\left(t\right)\geq 0 \\\eta_m\left(T_m\left(t\right),w_m\left(t\right)\right)\cdot w_m\left(t\right)\cdot T_m\left(t\right), ~T_m\left(t\right)<0
	\end{cases}\label{eq:ev-eco-drive-power-model}
\end{eqnarray}
where $\eta_m(\cdot)$ denotes the motor's efficiency map shown in Fig.~\ref{fig:ev-eco-operating-point}.




\subsection{Optimization Constraints}

\subsubsection{Traffic Constraints}

To promote safety and maintain efficient traffic flow, constraints are imposed on the following distance between the ego vehicle and its immediate preceding one.
Furthermore, prediction uncertainty is taken into account by integrating speed variance estimates obtained from the prediction algorithm shown in the previous section:
\begin{subequations}
\begin{align}
    d(t) &\geq d_\text{lead}(t) +\beta \sigma[d_\text{lead}(t)] - d_{max}\\
    d(t) &\leq d_\text{lead}(t) - \beta \sigma[d_\text{lead}(t)] - (d_{min} + h_{min}v(t))
\end{align}\label{eq:ev-eco-car-following-constraint}\end{subequations}  
\noindent
where $d_\text{lead}(t)$ represents the predicted longitudinal location of the vehicle directly ahead;
$d_{max}$ shows the maximum spacing, which is defined as \SI{40}{\meter} in this work;
$d_{min}$ represents the minimum spacing, which is set to \SI{5}{\meter};
$h_{min}$ denotes the time headway with value \SI{1.5}{\second};
the terms $\sigma$ and $\beta$ correspond to the standard deviation of the prediction and the associated confidence level, respectively.

\subsubsection{Sytem Constraints}
The vehicle's longitudinal speed must remain non-negative and below the speed limit $v_\text{lim}$.
Moreover, the acceleration is constrained by the tire's friction factor to ensure safe operation.
To maintain smooth and practical driving behavior, the rate of change of acceleration (jerk) is also limited in \eqref{eq:jerk}. Accordingly, we define:

\begin{subequations}
    \begin{align}
        0 \leq & v(t) \leq v_{\text{lim}},  \\
        \quad a_{{min}} \leq & a(t) \leq a_{{max}}   \\
        {j}_{{min}} \leq & \dot{a}(t) \leq {j}_{{max}} \label{eq:jerk}
    \end{align}
\end{subequations}

In addition, the maximum charging and discharging current, $I_\text{chg}(SOC(t))$ and $I_\text{dis}(SOC(t))$, typically vary with the battery’s $SOC$ and are subject to the constraint expressed by:

\begin{equation}
    I_\text{chg}(SOC(t)) \leq I_b(t) \leq I_\text{dis}(SOC(t))
    \label{eq:ev-eco-current-constraint}
\end{equation}

The electric motor's minimum and maximum torque outputs, ${T_m}_{min}(w_m(t))$ and ${T_m}_{max}(w_m(t))$ , respectively, are both functions of the motor rotating speed $w_m(t)$:

\begin{equation}
   {T_m}_{min}(w_m(t)) \leq T_m(t) \leq {T_m}_{max}(w_m(t)) 
\end{equation}

\subsubsection{Regenerative Braking Constraints}
In EVs, regenerative braking works together with the mechanical braking force $F_b(t)$ to slow down the vehicle.
To maximize energy efficiency, regenerative braking is ideally prioritized over mechanical braking, as it recovers kinetic energy, recharges the battery, and extends driving range.
However, the regenerative braking process is subject to various limitations. The proposed eco-driving controller optimally balances mechanical and regenerative braking forces while minimizing energy consumption and maintaining effective vehicle control.

Since the battery's maximum charging current is determined by its current $SOC$ level, the maximum regenerative braking force is also constrained.
This constraint is expressed in \eqref{eq:ev-eco-current-constraint} presented in the previous subsection.

{Meanwhile, to account for the electric motor's reduced efficiency when operating as a generator at low speeds, this study incorporates a low-speed cutoff point, defined as $v_\text{lscp} =$\SI{10}{\km/\hour} ~\cite{heydari2019maximizing}}:
\begin{equation}
    0 \leq T_m(t) \quad \text{if} \quad v(t) \leq v_\text{lscp}
    \label{eq:prediction-lscp}
\end{equation}

Furthermore, the following inequality is used to limit the maximum regenerative torque~\cite{ehsani2018modern}:
\begin{equation}\label{eq:ev-eco-constraint-max-reg}
    m\cdot a_\text{reg} \leq T_m(t) r_g(t)  
\end{equation}
where $a_\text{reg}$ represents the maximum deceleration achievable through regenerative braking, and is set to -0.2$g$. 
When integrated with an anti-lock braking systems (ABS) controller, this constraint can be formulated more strictly to ensure safety and stability.

\subsection{Design of Gear Ratios Selection}
To ensure the efficiency of the EV's powertrain system while maintaining its drivability, the following constraints are applied to determine the maximum and minimum lumped gear ratios $r_g(t)$ for the target EV's multi-speed transmission \cite{gao2015gear, ruan2016comparative, han2019optimized}:
\begin{subequations}
    \begin{align}
        T_{m_{max}} r_{max} \eta_\text{gear} &\geq {{f}_{\varphi}}_{max} \label{eq:ev-eco-max-gear-ratio}\\
        w_{m_{max}} / r_{min} &\geq {v}_{max}\label{eq:ev-eco-min-gear-ratio}
    \end{align}
    \label{eq:gear-ratio-design}
\end{subequations}
where $r_{max}$ and $r_{min}$ represent the maximum and minimum lumped gear ratios, respectively; ${f_{\varphi}}_{max}$ denotes the maximum rolling resistance caused by the road's slope angle.
\eqref{eq:ev-eco-max-gear-ratio} is used to ensure the vehicle's drivability when it drives uphills, 
while \eqref{eq:ev-eco-min-gear-ratio} is considered to guarantee the vehicle's ability to achieve its maximum speed.

\section{Problem Simplification} \label{sec:ev-eco-problem-simplification}
For EVs with multi-speed transmissions, the lumped ratio $r_g(t)$ in the powertrain model \eqref{eq:ev-eco-powertrain-model} and vehicle model \eqref{eq:ev-eco-vehicle-model} is a discrete input variable, resulting in a nonlinear mixed integer optimization problem.
However, this hybrid optimization problem is highly nonlinear and computationally intensive, rendering it inefficient to solve directly. 
To address this challenge and enable real-time computation, this problem is simplified in this section.

\subsection{Powertrain Model}

The powertrain model described in \eqref{eq:ev-eco-powertrain-model} exhibits bilinearity, involving both discrete and continuous variables.
To address this complexity, the big-M technique described in~\cite{bemporad1999control} is employed, converting this equality constraint into a set of linear inequality constraints.
For an EV transmission with $d_g$ gears and lumped ratios $r_{1}$,$r_{2}$, ..., $r_{d_g}$, a binary variable $g_i(t)\in\{0,1\}, i=1,2,..,d_g$, is introduced to indicate whether a specific gear is engaged, subject to the condition $\sum_{i=1}^{d_g}g_i(t) = 1$. 
Using these definitions, the constraint \eqref{eq:ev-eco-powertrain-model} can be reformulated as:
\begin{subequations}\label{eq:ev-eco-simplify-powertrain-model}
\begin{align}
    w_m(t) + (1-g_i(t))\cdot(r_i v_{max})&\geq r_i v(t)\\
    w_m(t) + (1-g_i(t))\cdot(-w_{m_{ub}})&\leq r_i v(t)    
\end{align}
\end{subequations}
where $v_{max}$ and $w_{ub}$ denote the upper limit of the vehicle driving speed and motor rotating speed, respectively.

The model described in \eqref{eq:ev-eco-vehicle-model} also includes the bilinear term $T_m(t)r_g(t)$.
By expressing the sum on its right-hand side as the resistance force $F_r(t)$, the equation can be reformulated using the big-M technique as:
\begin{subequations}
\label{eq:model-2}
    \begin{align}
        (1-g_i(t))(F_{r_{max}}+T_{m_{max}}r_i) )+T_m(t)r_i\geq F_r(t)\\
        (1-g_i(t))(F_{r_{min}}-T_{m_{max}}r_i)+ T_m(t) r_i\leq F_r(t)
    \end{align}
\end{subequations}
where $i=1, 2, ,...,d_g$;
$F_{r_{min}}$ and $F_{r_{max}}$ represent the lower and upper limits of resistance force, respectively.
The term $T_{m_{max}}$ indicates the upper bound of the motor torque.
{By using these two equations \eqref{eq:ev-eco-simplify-powertrain-model} and \eqref{eq:model-2}, the discrete gear selection using binary variables model is reformulated via the Big-M method into linear inequalities. This ensures that the mixed-integer co-optimization remains computationally feasible for real-time applications}

The resistance force $F_r(t)$ includes the quadratic term $v^2(t)$.
To improve computational efficiency, this term can be approximated using piecewise linear functions constructed with special ordered sets of type 2 (SOS2)~\cite{beale1976global}.
This approach allows for computational efficiency in solving the problem.
In this formulation, a set of non-negative variables variables ($\lambda_1, \lambda_2, ..., \lambda_{d_v}$) qualifies as SOS2 if no more than two variables are non-zero, and those two must be adjacent in the sequence.
{Given a set of $d_v=33$ increasing speed values ($v_1, v_2, v_3, .., v_{d_v}$) between \SI{0}{\meter/\second} and $v_\text{lim}=\SI{32}{\meter/\second}$, the speed $v(t)$ is represented as $v(t)=\sum_{i=1}^{d_v}\lambda_i(t)v_i$. The quadratic term $v^2(t)$ can then be approximated using the following constraints}:
\begin{equation}
v^2(t)=\sum_{i=1}^{d_v}\lambda_i(t)v^2_i
\end{equation}
with
\begin{equation}
~\sum_{i=1}^{d_v}\lambda_i(t) = 1;~\lambda_i\geq0,~\lambda_i(t)\text{ is SOS2,~}\forall i=1,2,...,d_v
\end{equation}

For the rolling resistance model ${f}_{\varphi}(t)$, since the road's slope angle will not change abruptly along the prediction horizon, we assume that this value remains constant throughout the prediction horizon. This value is taken to be the same as its initial value at the start of the optimization cycle.

\subsection{Cost Function}
\begin{figure}
    \setlength{\abovecaptionskip}{0.2cm}
    \centering
    \includegraphics[width=0.91\linewidth]{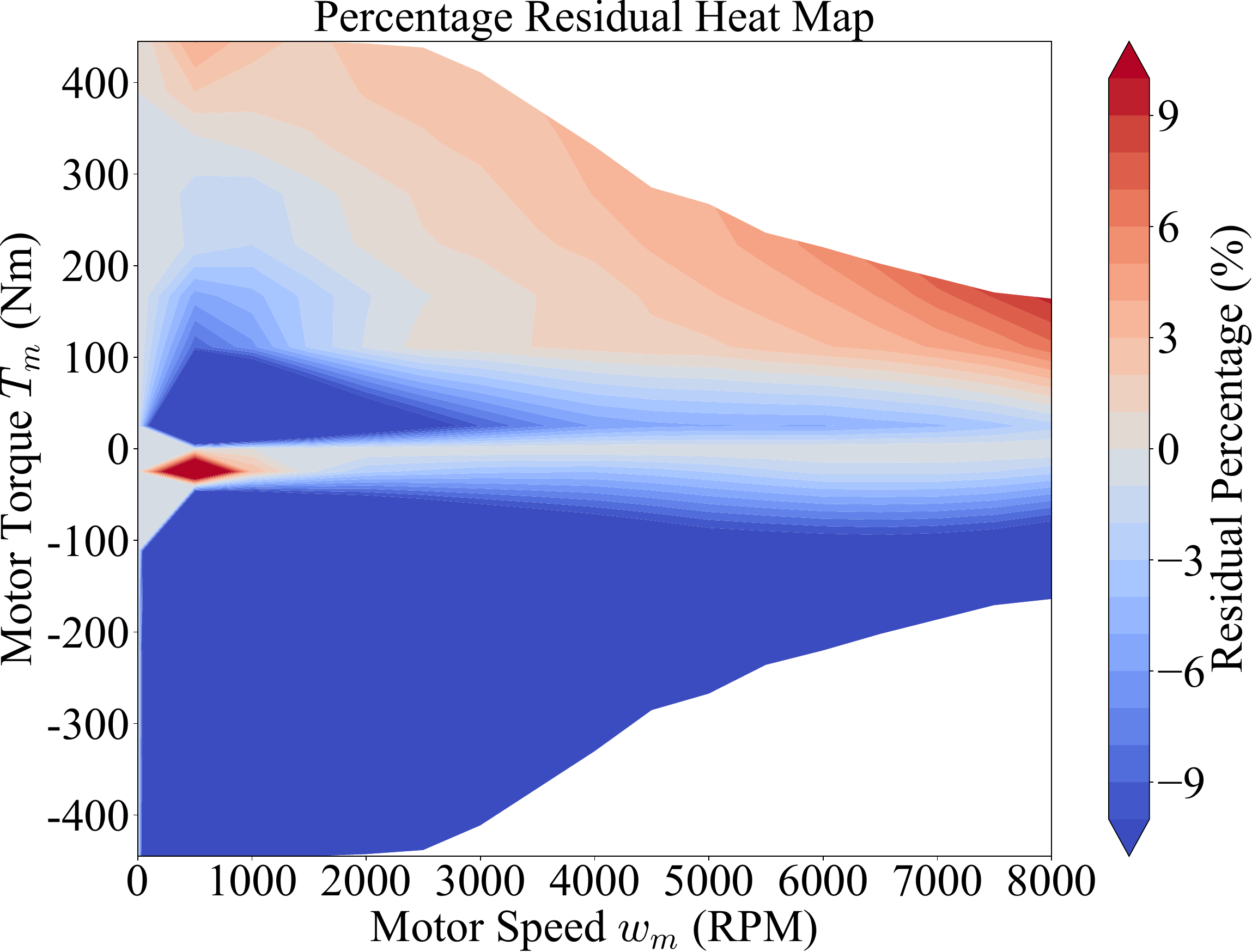}
    \caption{Residual heat map for motor power approximation.}
    \label{fig:motor-power-residual-map}
\end{figure}

The battery power $P_b(t)$ in the objective function \eqref{eq:ev-eco-objective-function} includes the drive power $P_\text{drv}(t)$, which is described by the nonlinear piecewise function~\eqref{eq:ev-eco-drive-power-model}.
This function represents the nonlinear motor efficiency map shown in Fig.~\ref{fig:ev-eco-operating-point}, posing significant computational challenges in optimization problems.
Therefore, to reduce the computational burden, $P_b(t)$ can be approximated as a polynomial function of $w_m(t)$ and $T_m(t)$:
\begin{eqnarray}
&~&\hskip-20pt P_{drv}(t) \approx p_{00}  + p_{10}\cdot w_m(t) + p_{01} \cdot T_m(t) \nonumber\\
    &~&\hskip 20pt + p_{11}\cdot w_m(t) \cdot T_m(t) \label{eq:ev-eco-p-drive-polynomial},
\end{eqnarray}
where $p_{ij}$ are fitting parameters that can be pre-calculated offline.
{Using the EV model provided by \cite{moswd2020autonomie}, the fitting parameters listed in Table \ref{tab:fitting-parameter-power-model} yield a coefficient of determination $R^2$ of 0.983 and a median percent error of 6.52\%, indicating excellent model accuracy.
The residual heat map for this approximation is shown in Fig.~\ref{fig:motor-power-residual-map}.}

\begin{table}[!htbp]
\centering
\begin{tabular}{c c c c}
  \hline
  $p_{00}$ & $p_{10}$ & $p_{01}$ & $p_{11}$ \\ \hline
  1344.5   & 1.64     & 28.1     & 1.0      \\ \hline
\end{tabular}
\caption{Fitting parameters for the drive power model $P_\text{drv}(t)$}
\label{tab:fitting-parameter-power-model}
\end{table}

However, since this approximation also includes the bilinear term $w_{m}(t)\cdot T_m(t)$, an auxiliary variable $P_m(t) = w_{m}(t)T_m(t)$ is introduced to simply the expression further.
The McCormick relaxation \cite{liberti2006exact} is used to convexify this bilinear term by converting them into a set of linear inequalities:

\begin{subequations}
    \begin{align}
        T_{m_{min}}w_m(t) + w_{m_{min}}T_m(t) - w_{m_{min}}T_{m_{min}}&\leq P_m(t)\\
        T_{m_{max}}w_m(t) + w_{m_{max}}T_m(t) - w_{m_{max}}T_{m_{max}}&\leq P_m(t)\\
        T_{m_{max}}w_m(t) + w_{m_{min}}T_m(t) - w_{m_{min}}T_{m_{max}}&\geq P_m(t)\\
        T_{m_{min}}w_m(t) + w_{m_{max}}T_m(t) - w_{m_{max}}T_{m_{min}}&\geq P_m(t)
    \end{align}
\end{subequations}
where $T_{m_{min}}$ and $T_{m_{max}}$ are the lower and upper bounds of motor torque; $w_{m_{min}}$ and $w_{m_{max}}$ are those of the motor rotating speed.

\subsection{Regenerative Braking}

\begin{figure*}
    \setlength{\abovecaptionskip}{0.2cm}
    \centering
    \includegraphics[width=0.9\linewidth]{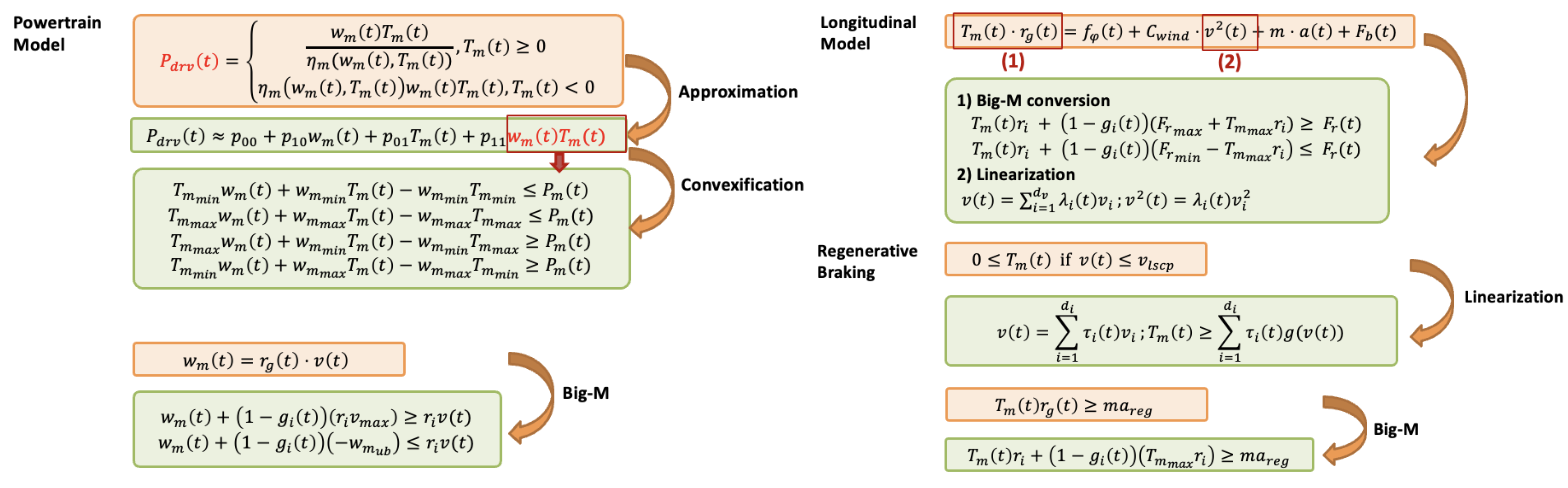}
    \caption{Schematic of the eco-driving co-optimization algorithm.}
    \label{fig:control flow}
\end{figure*}

For the constraint on low-speed cutoff point for regenerative braking, a step function $g(v(t))$ is used to represent this limitation.
{Similarly, SOS2 variables are used to approximate this nonlinear constraint with a piecewise linear representation.
Let $\tau_i(t)$ be the set of SOS2 variables and let ($v_{1}, v_{2},...,v_{d_l}$) be a sequence of $d_l=33$ equally spaced speed breaking points ranging from 0 to 32. The speed $v(t)$ is then approximated as $v(t)=\sum_{i=1}^{d_l}\tau_i(t)v_{i}$. To enforce this approximation, the following constraint is used}:
\begin{equation}
    T_m(t)\geq \sum_{i=1}^{d_l}\tau_i(t)g(v(t))
\end{equation}
with
\begin{equation}
    \sum_{i=1}^{d_l}\tau_i(t) = 1;~\tau_i\geq0,~\tau_i(t)\text{ is SOS2 }\forall i=1,2,...,d_l
\end{equation}


To address the bilinear term in the motor's maximum regeneration constraint~\eqref{eq:ev-eco-constraint-max-reg}, the big-M technique is also applied, resulting in the following linear approximation of the constraint:
\begin{equation}\label{eq:ev-eco-simplify-reg-constraint}
    T_m(t)r_i + (1-g_i(t))( T_{m_{max}}r_i)\geq m a_\text{reg} 
\end{equation}

\subsection{Optimization Constraints}

To speed up the computational process and ensure the feasibility of the MPC optimization problem, slack variables $s_{max}(t)$ and $s_{min}(t)$ are introduced into the car-following constraints \eqref{eq:ev-eco-car-following-constraint}:
\begin{subequations}
    \begin{align}
        d(t)&\geq d_\text{lead}(t) +\beta \sigma[d_\text{lead}(t)] - d_{max} - s_{max}(t)\\
        d(t)&\leq d_\text{lead}(t) -\beta \sigma[d_\text{lead}(t)]    - (d_{min}+h_{min}v(t))- s_{min}(t)
    \end{align}
\end{subequations}
with
\begin{equation}
    s_{max}(t)\geq0, ~s_{min}(t)\geq0
\end{equation}
Here, $s_{max}(t)$ and $s_{min}(t)$ are slack variables that transform hard constraints into cost-based soft constraints.
The term $w_{max} s^2_{max}(t) + w_{min} s^2_{min}(t) $, with two large positive weighting factors $w_{max}$ and $w_{min}$, is added to the objective function.

Since the battery's output voltage and internal resistance are assumed to be unchanged during one optimization cycle (with the battery $SOC$ changing insignificantly over 10--15 s), a constant relationship exists between the battery output power $P_b(t)$ and (dis)charging current $I_b(t)$.
To simplify the computational process, the battery output current $I_b(t)$ is not treated as an optimization variable and is instead directly calculated using $\eqref{eq:ev-eco-battery-model}$.
As a result, the constraint on the battery's maximum charging and discharging current in \eqref{eq:ev-eco-current-constraint} is reformulated as a constraint on the battery power output $P_b(t)$.

{Finally, the overall formulation and simplification of the eco-driving co-optimization algorithm are presented in Fig.~\ref{fig:control flow}.}

\subsection{Numerical Solution}\label{sec:ev-eco-numerical-solution}
In this work, the traffic prediction algorithm shown in Fig.~\ref{fig:ev-eco-prediction-structure} predicts the traffic states for a \SI{500}{\meter} road segment over the next \SI{10}{\second}, updating the prediction every second.
This road segment is divided into 20 equal-length cells--so that in \eqref{eq:ev-eco-pw} we have $dx=$\SI{25}{\meter}. 
The discretization time for this traffic prediction algorithm is set to $dt=$\SI{0.1}{\second}.
{To solve the mixed-integer optimization problem numerically, Gurobi \cite{optimization2020gurobi} is used to solve it in a MPC framework.}
The Euler method, with a discretization time of $dt=$\SI{0.2}{\second}, is employed to discrete the formulated optimization problem, enabling real-time computation and improving accuracy.
The prediction horizon for this MPC problem is set to \SI{10}{\second}, with the optimal solution updated every second.
{Python is used to implement the entire framework.
When executed on a computer equipped with an i7-8700 processor operating at 3.2GHz, Gurobi requires approximately \SI{0.27}{\second} on average to solve the simplified mixed-integer optimization problem in each cycle, which takes place every \SI{1}{\second}.}
These selected hyperparameters allow a balance between real-time capability and optimization performance.

\begin{figure*}
    \centering   
    \includegraphics[width=0.95\linewidth]{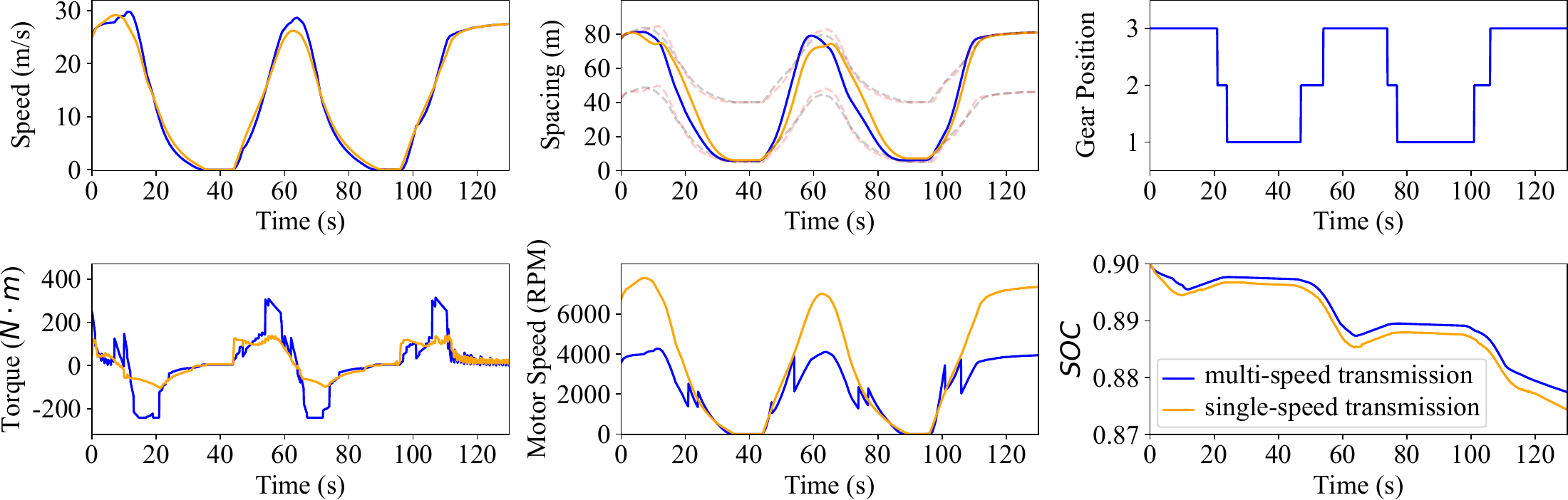}
    \caption{\textnormal{The comparison of the eco-driving co-optimization results for EVs equipped with three-speed and single-speed transmissions. Blue lines represent the states of the vehicle with three-speed transmission, while orange lines show those of the vehicle with single-speed transmission.
    In the spacing subplot, the red and black dashed lines correspond to the following distance constraints outlined in \eqref{eq:ev-eco-car-following-constraint}, representing the ego vehicle with three-speed and single-speed transmissions, respectively.}}
    \label{fig:ev-eco-optimization-compare}
\end{figure*}

\section{Results}\label{sec:ev-eco-results}

After presenting the proposed co-optimization based eco-driving controller for EVs with multi-speed transmissions, this section evaluates its performance using traffic scenarios from both numerical simulations and real-world road tests.
The effectiveness of the controller is compared to that of a single-speed counterpart, highlighting its advantages under diverse driving conditions.
In this study, the multi-speed EV is modeled with a three-speed transmission configuration.

For baseline comparison, the single-speed transmission EV is controlled using the identical co-optimization framework, with the lumped gear ratio $r_g$ held constant in the powertrain model \eqref{eq:ev-eco-prediction-powertrain}.
The setting of the traffic prediction algorithm remains consistent across both algorithms.
To compute the optimal solution of the optimization problem, IPOPT~\cite{biegler2009large} is used to solve it as a nonlinear MPC (NMPC) problem.
The Euler method, with a step length of $dt=$\SI{0.2}{\second}, is used to {discretize} the optimization problem, ensuring real-time computation and improving accuracy.
The prediction horizon for this MPC problem is set to \SI{10}{\second}, with the optimal solution {updated} every second.


Simulated traffic scenarios were created using the Simulation of Urban MObility (SUMO) tool~\cite{krajzewicz2002sumo}.
The simulations employed a time step of \SI{0.1}{\second}, with the Krauss model~\cite{krauss1998microscopic} selected as the car-following model. 
Road tests were conducted along a real-world signalized corridor on Trunk Highway 55 (TH55) in Minnesota, encompassing 22 intersections.
Roadside units (RSUs) installed at these intersections broadcast real-time SPaT information to onboard units (OBUs). The test vehicle recorded SPaT data, along with driving speed and location, to accurately capture real-world traffic conditions.

The second gear in the three-speed transmission shares the same gear ratio as the single-speed configuration, whereas the first and third gear ratios are specifically tailored to support the ego vehicle’s top speed and its ability to handle maximum road gradients.
Specifically, the lumped gear ratios $r_g(t)$ shown in \eqref{eq:ev-eco-powertrain-model} for the second gear is selected as 23.6 based on data sourced from \cite{moswd2020autonomie}.
Meanwhile, using \eqref{eq:gear-ratio-design}, {the lumped gear ratios for the first and third gears are selected as 17.1 and 32.3, respectively.}
{To consider the energy loss due to the introduction of multi-speed transmission, as discussed in \cite{ruan2017investigation}, the mechanical efficiency is set to 0.93 for the single-speed configuration and 0.83 for the three-speed setup.
These efficiency values are incorporated into the analysis.}
The vehicle weight $m$ with the single-speed transmission is \SI{1848}{\kilogram}.
{Since the two-speed transmission in production \cite{Drive2019Taycan} weighs around \SI{70}{\kilogram}, an extra \SI{100}{\kilogram} is added to the vehicle’s mass to account for the additional weight of the three-speed transmission.}
The drag coefficient $C_{wind}$ is 0.42.
Throughout this study, the ego vehicle is assumed to travel on flat terrain and the slope angle $\varphi(d(t))$ in \eqref{eq:ev-eco-powertrain-model} is consistently set to zero.

\subsection{Validation through Numerical Simulations}
In this subsection, the effectiveness of the developed controller design is validated using simulated traffic scenarios generated by SUMO.
The case study involves a 10-vehicle platoon, with the ego vehicle positioned at the rear.
The platoon travels along an \SI{1800}{\meter} road section over \SI{130}{\second}, encountering two signalized intersections and stopping at each due to red lights.
The scenario assumes that 50\% of the vehicles in the platoon are CVs, implying that the traffic prediction algorithm has access to speed and location data for only half of the vehicles.
To represent real-world scenarios, {whether a vehicle is a CV or non-CV is decided randomly in the simulation.}
The communication range for both V2V and V2I communications is set to \SI{500}{\meter}.

\begin{figure*}
    \setlength{\abovecaptionskip}{0.2cm}
    \centering
    \includegraphics[width=0.53\linewidth]{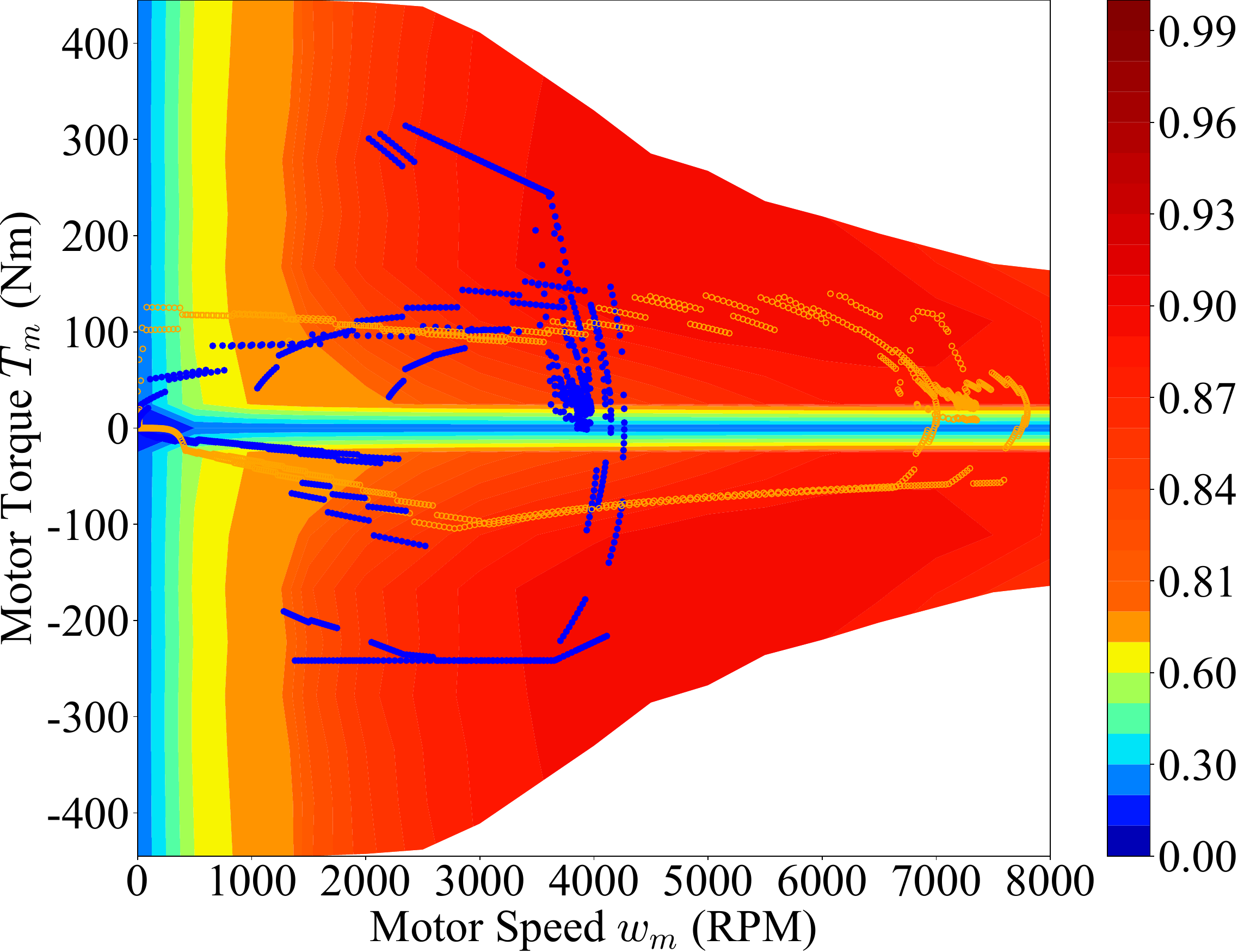}
    \caption{Comparison of motor operating points, where blue points indicate the operating conditions for the vehicle with a three-speed transmission, and orange points represent those for the single-speed transmission vehicle.}
    \label{fig:ev-eco-operating-point}
\end{figure*}

\begin{figure*}[!htbp]
    \centering   
    \includegraphics[width=0.95\linewidth]{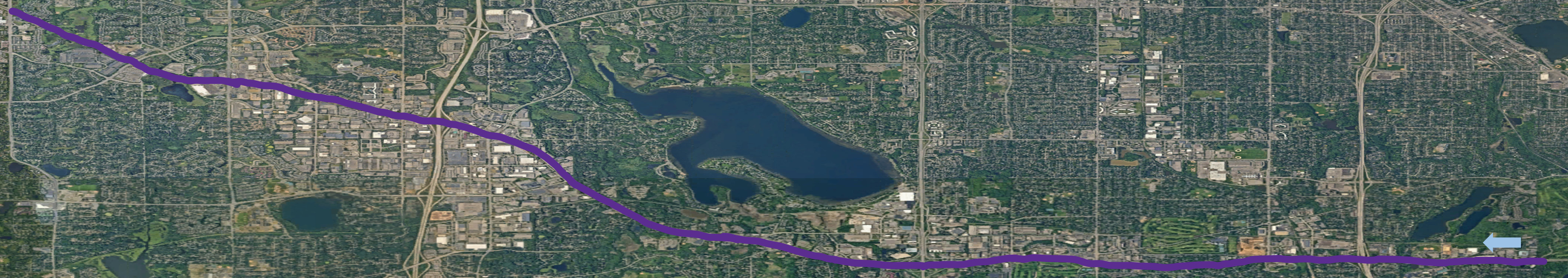}
    \caption{\textnormal{A map of Minnesota State Highway 55, with the road section used for the test drive highlighted in purple. The blue arrow indicates the direction of the test drive. }}
    \label{fig:ev-eco-th55-map}
\end{figure*}

In identical traffic scenarios, Fig.~\ref{fig:ev-eco-optimization-compare} illustrates the ego vehicle’s states for both configurations: the blue line represents the three-speed transmission, while the orange line corresponds to the single-speed transmission.
In the spacing subplot, the red and black dashed lines represent the limits on following distance shown in \eqref{eq:ev-eco-car-following-constraint} for ego vehicle with a three-speed transmission and a single-speed transmission, respectively.
Additionally, the motor operating points for these two vehicles are marked in Fig.~\ref{fig:ev-eco-operating-point} using blue and orange markers, respectively.

The results demonstrate that both vehicles effectively utilize the following distance range between the upper and lower spacing constraints, reflecting a typical eco-driving pattern that minimizes kinetic energy waste.
Due to the incorporation of the low-speed cutoff point constraint \eqref{eq:prediction-lscp} in the optimization framework, the motor torque for both vehicles remains non-negative in {low-speed} regions.
However, the introduction of a three-speed transmission allows the ego vehicle’s motor to operate consistently within a more efficient range, thereby reducing energy loss.
During acceleration, the three-speed transmission moves the motor’s operating points toward the optimal efficiency zone, in contrast to the single-speed transmission, which forces the motor to run at high torque and low speeds, demanding high current from the battery.
Similarly, regenerative braking becomes more effective.
At high speeds, the motor's reduced RPM with the three-speed transmission further minimizes energy losses.
Collectively, these factors result in significantly improved energy efficiency during both cruising and acceleration/deceleration phases, yielding substantial energy benefits.
{Then, the simulation is run for this \SI{1800}{\meter} road section another 9 times with randomly generated preceding vehicle behaviors to show the overall energy benefit of the proposed algorithm.
Over the combined simulation horizons, the energy efficiency, measured by battery $SOC$ consumption, improves by 12.01\% for the EV with a three-speed transmission, corresponding to the energy economy improving from 203.6 Wh/km to 179.1 Wh/km for the vehicle model used.}


\begin{figure*}[!htbp]
    \centering   
    \includegraphics[width=0.96\linewidth]{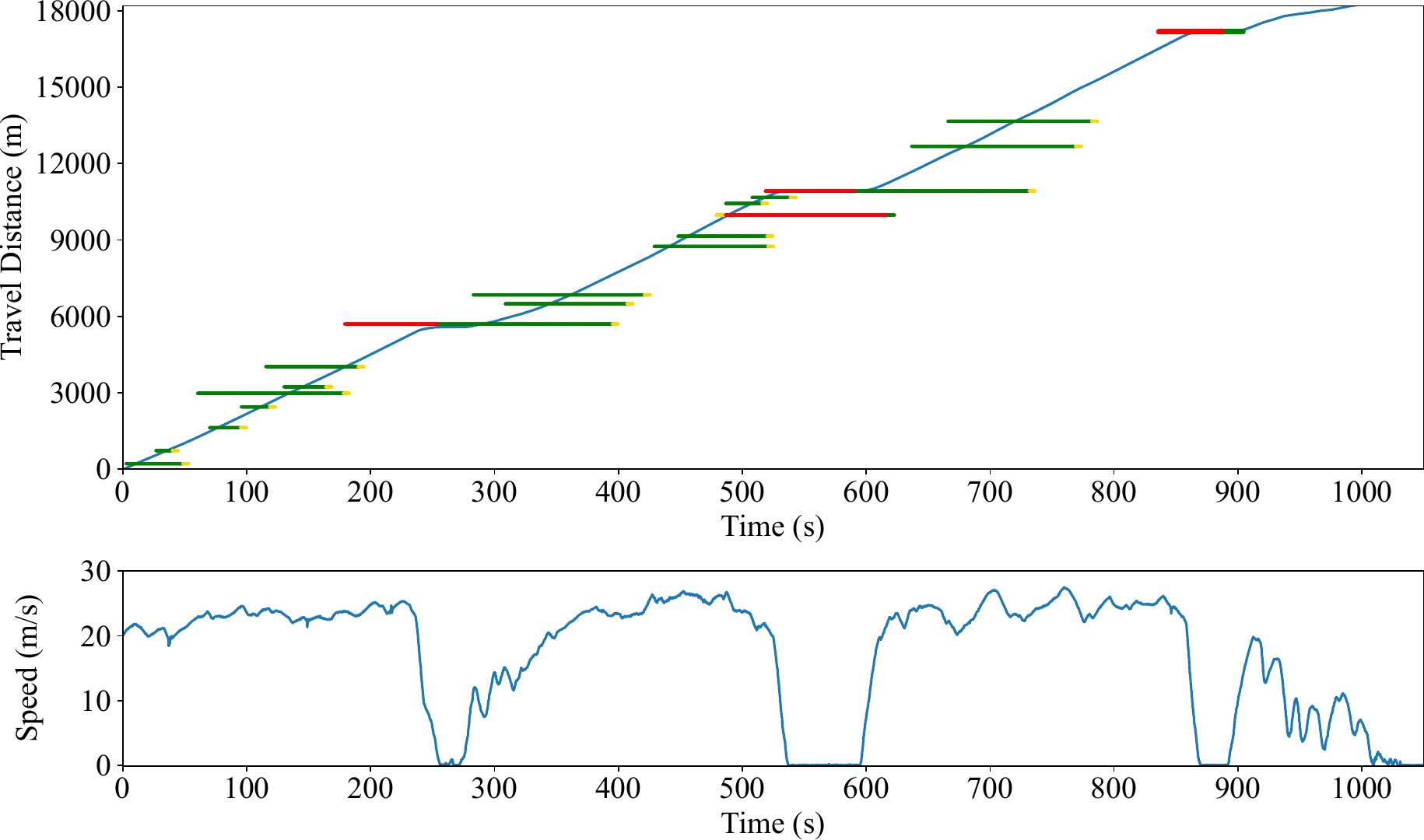}
    \caption{\textnormal{The test vehicle's longitudinal trajectory and speed profiles during a single test drive. The traffic signal status and its location are also shown in the first figure, reflecting when the traffic lights are within the test vehicle's communication range.}}
    \label{fig:ev-eco-th55-traj}
\end{figure*}

\begin{figure*}[!htbp]
    \centering   
    \includegraphics[width=0.96\linewidth]{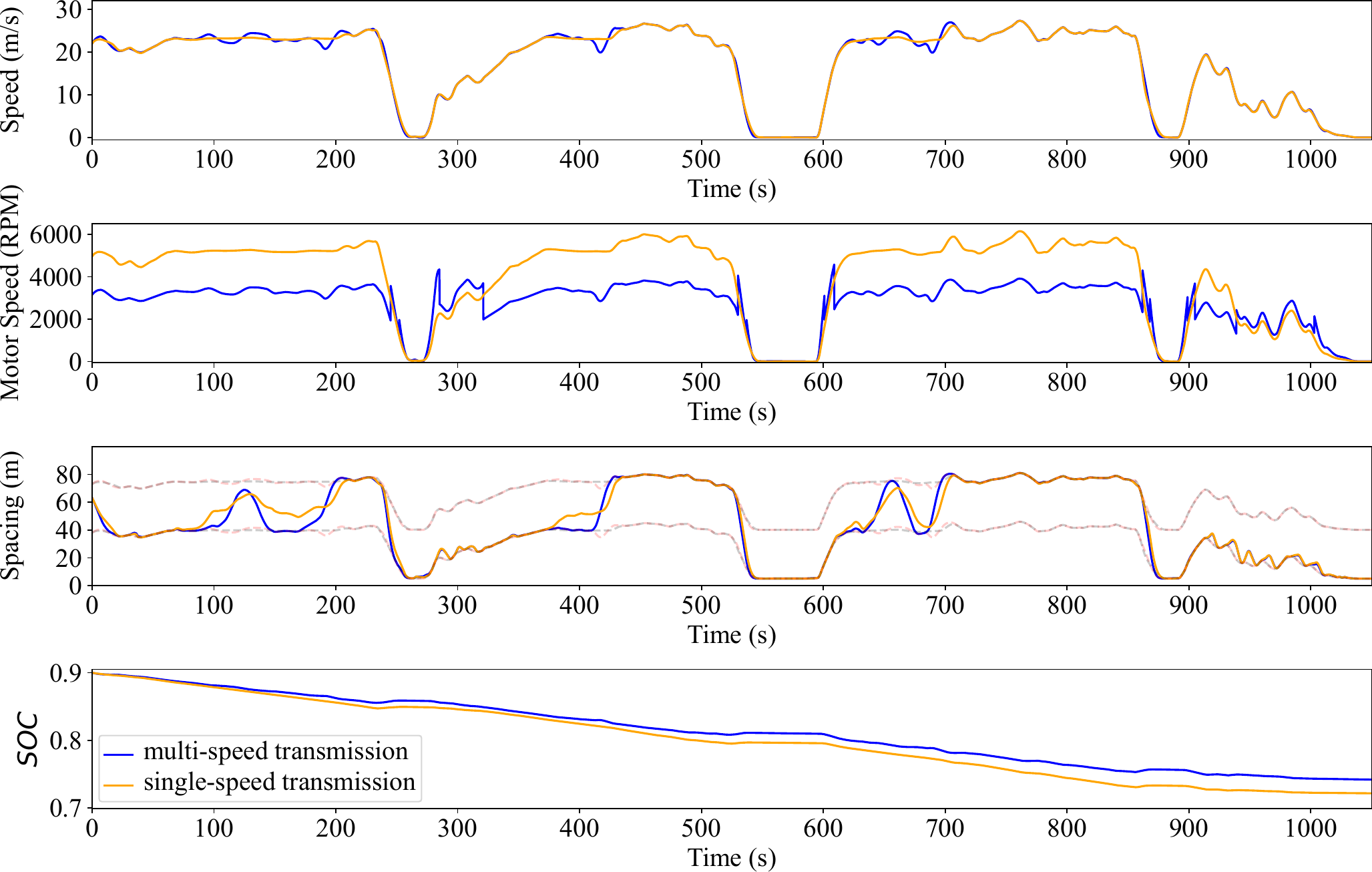}
    \caption{\textnormal{The comparison of the eco-driving co-optimization for EVs with three-speed and single-speed transmissions.
    The blue line represents the vehicle states for the three-speed configuration, while the orange line corresponds to the single-speed setup.
    In the spacing subplot, the red and black dashed lines indicate the following distance constraints of the ego vehicle with three-speed and single-speed transmissions, respectively.}}
    \label{fig:ev-eco-th55-optimization}
\end{figure*}
\subsection{Validation through Experimental Data}

To further validate our proposed eco-driving controller for EVs with multi-speed transmissions, we examine the developed controller using traffic data from real-world road tests.
These tests were conducted on Trunk Highway 55 (TH 55) in Minnesota (Fig.~\ref{fig:ev-eco-th55-map}) \cite{levin2023cost}.

Three test vehicles were equipped with OBUs to obtain real-time SPaT information, as well as their driving speeds and locations.
All aforementioned traffic information was recorded during the test.
These vehicles drove in the same lane, interleaved with several human-driven vehicles whose information was unavailable.
Figure \ref{fig:ev-eco-th55-traj} presents the trajectory, speed profiles, traffic-signal status (when the corresponding RSU was in range), and location data from the third test vehicle during a representative drive.
During this \SI{18}{\minute} test drive, test vehicles covered approximately \SI{18}{\km} and stopped three times at red signals.
To validate our eco-driving controller, we use this vehicle’s recorded data to simulate the immediate preceding vehicle in the eco-driving simulation, allowing the ego vehicle—running the eco-driving algorithm—to follow at a safe spacing.
The traffic prediction algorithm described in Sec. \ref{sec:ev-eco-background} was used to predict the trajectory of this third test vehicle based on data collected during the road test.

The traffic prediction algorithm depicted in Fig.~\ref{fig:ev-eco-prediction-structure} is then employed to forecast the longitudinal movement of the preceding vehicle throughout the entire process.
This trajectory serves as the traffic constraints \eqref{eq:ev-eco-car-following-constraint} in the co-optimization problem of the eco-driving controller.
Finally, Fig.~\ref{fig:ev-eco-th55-optimization} presents a comparison of the eco-driving controller’s performance between vehicles equipped with three-speed and single-speed transmissions, indicated by the blue and orange lines, respectively.

Consistent with the results using simulated traffic scenarios, our proposed controller for the EV with a three-speed transmission reduces energy consumption, in terms of battery $SOC$ consumption, by about 11.36\%  during this \SI{18}{\km} trip.
This improvement results from running the motor in a more efficient operating range than that of the single-speed transmission controller.
This further shows the importance of multi-speed transmissions in enhancing energy efficiency for electric vehicles in real-world scenarios.




\section{Conclusion} \label{sec:ev-eco-conclusion}
In this work, we show a real-time implementable eco-driving controller for EVs with multi-speed transmissions that optimizes vehicle speed and powertrain operation simultaneously.
We introduce the formulation of the co-optimization problem along with an approach to efficiently solve it through problem simplification.
We evaluate the developed algorithm via traffic scenarios from both numerical simulations and real-world road tests.
It is shown that the developed algorithm achieves a 12.01\% reduction in terms of battery $SOC$ usage during an \SI{1800}{\meter} drive in simulated traffic scenarios compared to that of eco-driving control for EV with single-speed transmission.
The traffic scenarios from real-world road tests show an 11.36\% energy consumption benefit over an \SI{18}{\km} drive.

{Although the proposed algorithm achieves a significant energy benefit for connected and autonomous electric vehicles with multi-speed transmissions, the primary limitation of this framework is the exclusion of battery thermal effects. Since temperature impacts both efficiency and degradation, integrating these thermal dynamics remains a key next step. This will expand our real-time controller into a more holistic powertrain management system, building upon the foundation of this work.}

\bibliographystyle{IEEEtran}
\bibliography{sampleBib}

@article{gao2025multi,
  title={Multi-objective co-optimization of powertrain sizing, energy management, and eco-driving for the architectural design of electric vehicles},
  author={Gao, Ye and Feng, Lei and Liu, Ding and Li, Zhiwu},
  journal={Applied Energy},
  volume={398},
  pages={126402},
  year={2025},
  publisher={Elsevier}
}

@article{lee2020model,
  title={Model-based reinforcement learning for eco-driving control of electric vehicles},
  author={Lee, Heeyun and Kim, Namwook and Cha, Suk Won},
  journal={IEEE Access},
  volume={8},
  pages={202886--202896},
  year={2020},
  publisher={IEEE}
}

@misc{Drive2019Taycan,
  title        = {Porsche Taycan: A look at its motors, transmission and dynamic chassis},
  organization = {Drive.com.au},
  howpublished = {\url{https://www.drive.com.au/news/} \url{porsche-taycan-a-look-at-its-motors-transmission-and-dynamic}
                 \url{-chassis}},
  note         = {Accessed: 2025-11-02}
}

@article{he2025connectivity,
  title={A Connectivity-Based Real-Time Traffic Prediction Considering Lane-Changing Maneuvers with Application to Eco-Driving Control of Electric Vehicles},
  author={He, Suiyi and Wang, Shian and Shao, Yunli and Sun, Zongxuan and Levin, Michael W},
  journal={IEEE Transactions on Vehicular Technology},
  year={2025},
  publisher={IEEE}
}

@inproceedings{sun2021traffic,
  title={Traffic Prediction for Connected Vehicles on a Signalized Arterial},
  author={Sun, Wenbo and Wang, Shian and Shao, Yunli and Sun, Zongxuan and Levin, Michael W},
  booktitle={2021 IEEE International Intelligent Transportation Systems Conference},
  pages={1968--1973},
  year={2021},
}

@article{sun2022energy,
  title={Energy and mobility impacts of connected autonomous vehicles with co-optimization of speed and powertrain on mixed vehicle platoons},
  author={Sun, Wenbo and Wang, Shian and Shao, Yunli and Sun, Zongxuan and Levin, Michael W},
  journal={Transportation Research Part C: Emerging Technologies},
  volume={142},
  pages={103764},
  year={2022},
  publisher={Elsevier}
}

@article{shao2020eco,
  title={Eco-approach with traffic prediction and experimental validation for connected and autonomous vehicles},
  author={Shao, Yunli and Sun, Zongxuan},
  journal={IEEE Transactions on Intelligent Transportation Systems},
  volume={22},
  number={3},
  pages={1562--1572},
  year={2021},
}

@article{shao2020vehicle,
  title={Vehicle speed and gear position co-optimization for energy-efficient connected and autonomous vehicles},
  author={Shao, Yunli and Sun, Zongxuan},
  journal={IEEE Transactions on Control Systems Technology},
  volume={29},
  number={4},
  pages={1721--1732},
  year={2021},
}

@inproceedings{han2019optimized,
  title={Optimized design of multi-speed transmissions for battery electric vehicles},
  author={Han, Kyoungseok and Wang, Yan and Filev, Dimitar and Dai, Edward and Kolmanovsky, Ilya and Girard, Anouck},
  booktitle={2019 American Control Conference (ACC)},
  pages={816--821},
  year={2019},
  organization={IEEE}
}

@article{huang2020optimal,
  title={Optimal design and control of a two-speed planetary gear automatic transmission for electric vehicle},
  author={Huang, Wei and Huang, Jianfeng and Yin, Chengliang},
  journal={Applied sciences},
  volume={10},
  number={18},
  pages={6612},
  year={2020},
  publisher={MDPI}
}

@article{bemporad1999control,
  title={Control of systems integrating logic, dynamics, and constraints},
  author={Bemporad, Alberto and Morari, Manfred},
  journal={Automatica},
  volume={35},
  number={3},
  pages={407--427},
  year={1999},
  publisher={Elsevier}
}

@article{liberti2006exact,
  title={An exact reformulation algorithm for large nonconvex {NLP}s involving bilinear terms},
  author={Liberti, Leo and Pantelides, Constantinos C},
  journal={Journal of Global Optimization},
  volume={36},
  number={2},
  pages={161--189},
  year={2006},
  publisher={Springer}
}

@article{wang2005real,
  title={Real-time freeway traffic state estimation based on extended Kalman filter: a general approach},
  author={Wang, Yibing and Papageorgiou, Markos},
  journal={Transportation Research Part B: Methodological},
  volume={39},
  number={2},
  pages={141--167},
  year={2005},
  publisher={Elsevier}
}

@inproceedings{shao2019ASME,
  title={Optimal speed control for a connected and autonomous electric vehicle considering battery aging and regenerative braking limits},
  author={Shao, Yunli and Sun, Zongxuan},
  booktitle={ASME 2019 Dynamic Systems and Control Conference},
  year={2019},
  organization={American Society of Mechanical Engineers Digital Collection}
}

@article{shao2021energy,
  title={Energy-Efficient connected and automated vehicles: real-time traffic prediction-enabled co-optimization of vehicle motion and powertrain operation},
  author={Shao, Yunli and Sun, Zongxuan},
  journal={IEEE Vehicular Technology Magazine},
  volume={16},
  number={3},
  pages={47--56},
  year={2021},
}

@inproceedings{wan2000unscented,
  title={The unscented Kalman filter for nonlinear estimation},
  author={Wan, Eric A and Van Der Merwe, Rudolph},
  booktitle={IEEE Adaptive Systems for Signal Processing, Communications, and Control Symposium},
  pages={153--158},
  year={2000},
}

@inproceedings{krajzewicz2002sumo,
  title={{SUMO (Simulation of Urban MObility)}-an open-source traffic simulation},
  author={Krajzewicz, Daniel and Hertkorn, Georg and R{\"o}ssel, Christian and Wagner, Peter},
  booktitle={Proceedings of the 4th Middle East Symposium on Simulation and Modelling},
  pages={183--187},
  year={2002}
}

@techreport{moswd2020autonomie,
  title={{AUTONOMIE VID}},
  author={Moswd, Ayman and Kim, Namdoo and others},
  year={2020},
  institution={Argonne National Lab (ANL), Argonne, IL}
}

@inproceedings{shao2019optimal,
  title={Optimal speed control for a connected and autonomous electric vehicle considering battery aging and regenerative braking limits},
  author={Shao, Yunli and Sun, Zongxuan},
  booktitle={Dynamic Systems and Control Conference},
  volume={59148},
  year={2019},
  organization={American Society of Mechanical Engineers}
}

@article{biegler2009large,
  title={Large-scale nonlinear programming using IPOPT: An integrating framework for enterprise-wide dynamic optimization},
  author={Biegler, Lorenz T and Zavala, Victor M},
  journal={Computers \& Chemical Engineering},
  volume={33},
  number={3},
  pages={575--582},
  year={2009},
  publisher={Elsevier}
}

@article{koch2021eco,
  title={Eco-driving for different electric powertrain topologies considering motor efficiency},
  author={Koch, Alexander and B{\"u}rchner, Tim and Herrmann, Thomas and Lienkamp, Markus},
  journal={World Electric Vehicle Journal},
  volume={12},
  number={1},
  pages={6},
  year={2021},
  publisher={MDPI}
}

@inproceedings{zhao2025design,
  title={Design, Implementation, and Evaluation of an Innovative Vehicle-Powertrain Eco-Operation System for Plug-In Hybrid Electric Buses},
  author={Zhao, Zhouqiao and Wu, Guoyuan and Hao, Peng and Ye, Fei and Gao, Zhiming and LaClair, Tim J and Brown, Dylan and Esaid, Danial and Boriboonsomsin, Kanok and Barth, Matthew J},
  booktitle={2025 IEEE Conference on Technologies for Sustainability (SusTech)},
  pages={1--8},
  year={2025},
  organization={IEEE}
}

@article{mazali2022review,
  title={Review of the methods to optimize power flow in electric vehicle powertrains for efficiency and driving performance},
  author={Mazali, Izhari Izmi and Daud, Zul Hilmi Che and Hamid, Mohd Kameil Abdul and Tan, Victor and Samin, Pakharuddin Mohd and Jubair, Abdullah and Ibrahim, Khairul Amilin and Kob, Mohd Salman Che and Xinrui, Wang and Talib, Mat Hussin Ab},
  journal={Applied Sciences},
  volume={12},
  number={3},
  pages={1735},
  year={2022},
  publisher={MDPI}
}

@book{ehsani2018modern,
  title={Modern electric, hybrid electric, and fuel cell vehicles},
  author={Ehsani, Mehrdad and Gao, Yimin and Longo, Stefano and Ebrahimi, Kambiz M},
  year={2018},
  publisher={CRC Press}
}

@article{heydari2019maximizing,
  title={Maximizing regenerative braking energy recovery of electric vehicles through dynamic low-speed cutoff point detection},
  author={Heydari, Shoeib and Fajri, Poria and Rasheduzzaman, Md and Sabzehgar, Reza},
  journal={IEEE Transactions on Transportation Electrification},
  volume={5},
  number={1},
  pages={262--270},
  year={2019},
}

@article{ruan2017investigation,
  title={An investigation of hybrid energy storage system in multi-speed electric vehicle},
  author={Ruan, Jiageng and Walker, Paul David and Zhang, Nong and Wu, Jinglai},
  journal={Energy},
  volume={140},
  pages={291--306},
  year={2017},
  publisher={Elsevier}
}

@article{beale1976global,
  title={Global optimization using special ordered sets},
  author={Beale, EML and Forrest, John JH},
  journal={Mathematical Programming},
  volume={10},
  pages={52--69},
  year={1976},
  publisher={Springer}
}

@misc{optimization2020gurobi,
  title={Gurobi optimizer reference manual},
  author={Optimization, Gurobi and others},
  year={2020}
}

@article{liu2023study,
  title={Study on the design and speed ratio control strategy of continuously variable transmission for electric vehicle},
  author={Liu, Yunfeng and Liu, Kan and Zhou, Yunshan and Chen, Yongdan and Wei, Dong and Zhou, Shichao and Luan, Haozhe},
  journal={IEEE Access},
  volume={11},
  pages={107880--107891},
  year={2023},
  publisher={IEEE}
}

@article{ruan2016comparative,
  title={A comparative study energy consumption and costs of battery electric vehicle transmissions},
  author={Ruan, Jiageng and Walker, Paul and Zhang, Nong},
  journal={Applied energy},
  volume={165},
  pages={119--134},
  year={2016},
  publisher={Elsevier}
}

@article{gao2015gear,
  title={Gear ratio optimization and shift control of 2-speed I-AMT in electric vehicle},
  author={Gao, Bingzhao and Liang, Qiong and Xiang, Yu and Guo, Lulu and Chen, Hong},
  journal={Mechanical Systems and Signal Processing},
  volume={50},
  pages={615--631},
  year={2015},
  publisher={Elsevier}
}

@inproceedings{han2020hierarchical,
  title={Hierarchical optimization of speed and gearshift control for battery electric vehicles using preview information},
  author={Han, Kyoungseok and Li, Nan and Kolmanovsky, Ilya and Girard, Anouck and Wang, Yan and Filev, Dimitar and Dai, Edward},
  booktitle={2020 American Control Conference (ACC)},
  pages={4913--4919},
  year={2020},
  organization={IEEE}
}

@article{li2021coordinated,
  title={Coordinated receding-horizon control of battery electric vehicle speed and gearshift using relaxed mixed-integer nonlinear programming},
  author={Li, Nan and Han, Kyoungseok and Kolmanovsky, Ilya and Girard, Anouck},
  journal={IEEE Transactions on Control Systems Technology},
  volume={30},
  number={4},
  pages={1473--1483},
  year={2021},
  publisher={IEEE}
}

@article{gao2022topology,
  title={Topology optimization and the evolution trends of two-speed transmission of EVs},
  author={Gao, Bingzhao and Meng, Dele and Shi, Wentong and Cai, Wenqi and Dong, Shiying and Zhang, Yuanjian and Chen, Hong},
  journal={Renewable and Sustainable Energy Reviews},
  volume={161},
  pages={112390},
  year={2022},
  publisher={Elsevier}
}

@article{fang2016design,
  title={Design and control of a novel two-speed uninterrupted mechanical transmission for electric vehicles},
  author={Fang, Shengnan and Song, Jian and Song, Haijun and Tai, Yuzhuo and Li, Fei and Nguyen, Truong Sinh},
  journal={Mechanical Systems and Signal Processing},
  volume={75},
  pages={473--493},
  year={2016},
  publisher={Elsevier}
}

@article{li2021online,
  title={Online optimization of gear shift and velocity for eco-driving using adaptive dynamic programming},
  author={Li, Guoqiang and G{\"o}rges, Daniel and Wang, Meng},
  journal={IEEE Transactions on Intelligent Vehicles},
  volume={7},
  number={1},
  pages={123--132},
  year={2021},
  publisher={IEEE}
}

@article{guo2016online,
  title={Online shift schedule optimization of 2-speed electric vehicle using moving horizon strategy},
  author={Guo, Lulu and Gao, Bingzhao and Chen, Hong},
  journal={IEEE/ASME Transactions on Mechatronics},
  volume={21},
  number={6},
  pages={2858--2869},
  year={2016},
  publisher={IEEE}
}

@article{guo2017line,
  title={On-line optimal control of the gearshift command for multispeed electric vehicles},
  author={Guo, Lulu and Gao, Bingzhao and Liu, Qifang and Tang, Jiahui and Chen, Hong},
  journal={IEEE/ASME Transactions on Mechatronics},
  volume={22},
  number={4},
  pages={1519--1530},
  year={2017},
  publisher={IEEE}
}

@article{bentaleb2024gear,
  title={Gear Shifting and Vehicle Speed Optimization for Eco-Driving on Curved Roads},
  author={Bentaleb, Ahmed and El Hajjaji, Ahmed and Rabhi, Abdelhamid and Karama, Asma and Benzaouia, Abdellah},
  journal={IEEE Access},
  year={2024},
  publisher={IEEE}
}

@article{liao2021eco,
  title={An eco-driving strategy for electric vehicle based on the powertrain},
  author={Liao, Peng and Tang, Tie-Qiao and Liu, Ronghui and Huang, Hai-Jun},
  journal={Applied Energy},
  volume={302},
  pages={117583},
  year={2021},
  publisher={Elsevier}
}

@article{kim2024driving,
  title={Driving Control Strategy and Specification Optimization for All-Wheel-Drive Electric Vehicle System with a Two-Speed Transmission},
  author={Kim, Jeonghyuk and Ahn, Jihyeok and Jeong, Seyoung and Park, Young-Geun and Kim, Hyobin and Cho, Dongwook and Hwang, Sung-Ho},
  journal={World Electric Vehicle Journal},
  volume={15},
  number={10},
  pages={476},
  year={2024},
  publisher={MDPI}
}

@inproceedings{he2023real,
  title={Real-time traffic prediction considering lane changing maneuvers with application to eco-driving control of electric vehicles},
  author={He, Suiyi and Wang, Shian and Shao, Yunli and Sun, Zongxuan and Levin, Michael W},
  booktitle={2023 IEEE Intelligent Vehicles Symposium (IV)},
  pages={1--7},
  year={2023},
  organization={IEEE}
}

@techreport{saini2024innovating,
  title={Innovating Mobility: The Design and Optimization of an Efficient Two-Speed Transmission for EVs},
  author={Saini, Sandeep and Rodrigues, Keith and Jennings, John and Finn, Dustin},
  year={2024},
  institution={SAE Technical Paper}
}

@article{kwon2023optimization,
  title={Optimization of multi-speed transmission for electric vehicles based on electrical and mechanical efficiency analysis},
  author={Kwon, Kihan and Lee, Jung-Hwan and Lim, Sang-Kil},
  journal={Applied Energy},
  volume={342},
  pages={121203},
  year={2023},
  publisher={Elsevier}
}

@article{nguyen2021optimization,
  title={Optimization and coordinated control of gear shift and mode transition for a dual-motor electric vehicle},
  author={Nguyen, Cong Thanh and Walker, Paul D and Zhang, Nong},
  journal={Mechanical Systems and Signal Processing},
  volume={158},
  pages={107731},
  year={2021},
  publisher={Elsevier}
}

@article{zhou2025design,
  title={Design and Optimization of an Electric Vehicle Powertrain Based on an Electromechanical Efficiency Analysis},
  author={Zhou, Baoyu and Li, Zhejun and Wang, Haichang and Cui, Yunxiang and Hu, Jie and Jiang, Feng},
  journal={Processes},
  volume={13},
  number={6},
  pages={1698},
  year={2025},
  publisher={MDPI}
}

@article{tian2020optimal,
  title={Optimal coordinating gearshift control of a two-speed transmission for battery electric vehicles},
  author={Tian, Yang and Yang, Haitao and Mo, Wenwei and Zhou, Shilei and Zhang, Nong and Walker, Paul D},
  journal={Mechanical Systems and Signal Processing},
  volume={136},
  pages={106521},
  year={2020},
  publisher={Elsevier}
}

@article{sorniotti2012analysis,
  title={Analysis and simulation of the gearshift methodology for a novel two-speed transmission system for electric powertrains with a central motor},
  author={Sorniotti, Aldo and Holdstock, Thomas and Pilone, Gabriele Loro and Viotto, Fabio and Bertolotto, Stefano and Everitt, Mike and Barnes, Robert J and Stubbs, Ben and Westby, Matt},
  journal={Proceedings of the Institution of Mechanical Engineers, Part D: Journal of Automobile Engineering},
  volume={226},
  number={7},
  pages={915--929},
  year={2012},
  publisher={Sage Publications Sage UK: London, England}
}

@article{qi2017analysis,
  title={Analysis and optimization of the gear-shifting process for automated manual transmissions in electric vehicles},
  author={Qi, Xiaowei and Yang, Yiyong and Wang, Xiangyu and Zhu, Zaobei},
  journal={Proceedings of the Institution of Mechanical Engineers, Part D: Journal of Automobile Engineering},
  volume={231},
  number={13},
  pages={1751--1765},
  year={2017},
  publisher={SAGE Publications Sage UK: London, England}
}

@article{zhu2015gear,
  title={Gear shift schedule design for multi-speed pure electric vehicles},
  author={Zhu, Bo and Zhang, Nong and Walker, Paul and Zhou, Xingxing and Zhan, Wenzhang and Wei, Yueyuan and Ke, Nanji},
  journal={Proceedings of the Institution of Mechanical Engineers, Part D: Journal of Automobile Engineering},
  volume={229},
  number={1},
  pages={70--82},
  year={2015},
  publisher={SAGE Publications Sage UK: London, England}
}

@article{hong2016shift,
  title={Shift control of a dry-type two-speed dual-clutch transmission for an electric vehicle},
  author={Hong, Sungwha and Son, Hanho and Lee, Seulgi and Park, Jongyun and Kim, Kyungha and Kim, Hyunsoo},
  journal={Proceedings of the Institution of Mechanical Engineers, Part D: Journal of Automobile Engineering},
  volume={230},
  number={3},
  pages={308--321},
  year={2016},
  publisher={SAGE Publications Sage UK: London, England}
}

@article{lacock2023electric,
  title={Electric vehicle drivetrain efficiency and the multi-speed transmission question},
  author={Lacock, Stephan and du Plessis, Armand Andr{\'e} and Booysen, Marthinus Johannes},
  journal={World Electric Vehicle Journal},
  volume={14},
  number={12},
  pages={342},
  year={2023},
  publisher={MDPI}
}

@article{yang2023optimal,
  title={Optimal control for shifting command of two-speed electric vehicles considering shifting loss},
  author={Yang, Liyue and Park, Dohyun and Lyu, Shaowen and Zheng, Chunhua and Kim, Namwook},
  journal={International Journal of Automotive Technology},
  volume={24},
  number={4},
  pages={1051--1059},
  year={2023},
  publisher={Springer}
}

@article{chai2019compound,
  title={Compound optimal control for shift processes of a two-speed automatic mechanical transmission in electric vehicles},
  author={Chai, Benben and Zhang, Jianwu and Wu, Shaofang},
  journal={Proceedings of the Institution of Mechanical Engineers, Part D: Journal of Automobile Engineering},
  volume={233},
  number={8},
  pages={2213--2231},
  year={2019},
  publisher={SAGE Publications Sage UK: London, England}
}

@article{naeem2022eco,
  title={Eco-driving control of electric vehicle with battery dynamic model and multiple traffic signals},
  author={Naeem, Hafiz Muhammad Yasir and Bhatti, Aamer Iqbal and Butt, Yasir Awais and Ahmed, Qadeer},
  journal={Proceedings of the Institution of Mechanical Engineers, Part D: Journal of Automobile Engineering},
  volume={236},
  number={6},
  pages={1133--1143},
  year={2022},
  publisher={SAGE Publications Sage UK: London, England}
}

@article{levin2023cost,
  title={Cost/Benefit Analysis of Fuel-Efficient Speed Control Using Signal Phasing and Timing (SPaT) Data: Evaluation for Future Connected Corridor Deployment},
  author={Levin, Michael W and Sun, Zongxuan and Wang, Shi’an and Sun, Wenbo and He, Suiyi and Suh, Bohoon and Zhao, Gaonan and Margolis, Jacob and Zamanpour, Maziar},
  year={2023},
  publisher={Minnesota Department of Transportation}
}

@article{rabinowitz2023real,
  title={Real-time implementation comparison of urban eco-driving controls},
  author={Rabinowitz, Aaron I and Ang, Chon Chia and Mahmoud, Yara Hazem and Araghi, Farhang Motallebi and Meyer, Richard T and Kolmanovsky, Ilya and Asher, Zachary D and Bradley, Thomas H},
  journal={IEEE Transactions on Control Systems Technology},
  year={2023},
  publisher={IEEE}
}

@article{fan2024deep,
  title={Deep Reinforcement Learning Based Integrated Eco-driving Strategy for Connected and Automated Electric Vehicles in Complex Urban Scenarios},
  author={Fan, Jiawei and Li, Jie and Wu, Xiaodong},
  journal={IEEE Transactions on Vehicular Technology},
  year={2024},
  publisher={IEEE}
}

@article{sun2023coordinated,
  title={Coordinated Hierarchical Co-Optimization of Speed Planning and Energy Management for Electric Vehicles Driving in Stochastic Environment},
  author={Sun, Chao and Zhang, Chuntao and Zhou, Xingyu and Sun, Fengchun},
  journal={IEEE Transactions on Vehicular Technology},
  volume={72},
  number={10},
  pages={12628--12638},
  year={2023},
  publisher={IEEE}
}

@book{sun2014design,
  title={Design and control of automotive propulsion systems},
  author={Sun, Zongxuan and Zhu, Guoming G},
  year={2014},
  publisher={CRC press}
}

@misc{
energy2024,
author="EIA",
title="U.S. energy facts explained",
howpublished="Website",
year={2024},
note={\url{https://www.eia.gov/energyexplained/us-energy-facts/}}
}

@article{muratori2021rise,
  title={The rise of electric vehicles—2020 status and future expectations},
  author={Muratori, Matteo and others},
  journal={Progress in Energy},
  volume={3},
  number={2},
  pages={022002},
  year={2021},
  publisher={IOP Publishing}
}

@article{li2024review,
  title={Review on eco-driving control for connected and automated vehicles},
  author={Li, Jie and Fotouhi, Abbas and Liu, Yonggang and Zhang, Yuanjian and Chen, Zheng},
  journal={Renewable and Sustainable Energy Reviews},
  volume={189},
  pages={114025},
  year={2024},
  publisher={Elsevier}
}

@article{ahssan2018electric,
  title={Electric vehicle with multi-speed transmission: A review on performances and complexities},
  author={Ahssan, Md Ragib and Ektesabi, Mehran Motamed and Gorji, Saman Asghari},
  journal={SAE International Journal of Alternative Powertrains},
  volume={7},
  number={2},
  pages={169--182},
  year={2018},
  publisher={JSTOR}
}

@article{hamednia2023optimal,
  title={Optimal thermal management, charging, and eco-driving of battery electric vehicles},
  author={Hamednia, Ahad and Murgovski, Nikolce and Fredriksson, Jonas and Forsman, Jimmy and Pourabdollah, Mitra and Larsson, Viktor},
  journal={IEEE Transactions on Vehicular Technology},
  volume={72},
  number={6},
  pages={7265--7278},
  year={2023},
  publisher={IEEE}
}

@article{coppola2022eco,
  title={Eco-driving control architecture for platoons of uncertain heterogeneous nonlinear connected autonomous electric vehicles},
  author={Coppola, Angelo and Lui, Dario Giuseppe and Petrillo, Alberto and Santini, Stefania},
  journal={IEEE Transactions on Intelligent Transportation Systems},
  volume={23},
  number={12},
  pages={24220--24234},
  year={2022},
  publisher={IEEE}
}

@techreport{krauss1998microscopic,
  title={Microscopic modeling of traffic flow: Investigation of collision free vehicle dynamics},
  author={Krauss, Stefan},
  place = {Germany},
  year = {1998},
  month = {April}
}


\end{document}